\documentclass[12pt]{article}
\usepackage{amssymb,graphicx}

\begin{document}

\newcommand{\sect}[1]{\setcounter{equation}{0}\section{#1}}
\renewcommand{\theequation}{\thesection.\arabic{equation}}
\newcommand{\be}{\begin{equation}}
\newcommand{\ee}{\end{equation}}
\newcommand{\bea}{\begin{eqnarray}}
\newcommand{\eea}{\end{eqnarray}}
\newcommand{\nonu}{\nonumber\\}
\newcommand{\beano}{\begin{eqnarray*}}
\newcommand{\eeano}{\end{eqnarray*}}
%%%%%%%%%%%%%%%%%%%%%%%%%
%%%%%%%%     GREQUES     %%%%%%%
%%%%%%%%%%%%%%%%%%%%%%%%%
\newcommand{\eps}{\epsilon}
\newcommand{\om}{\omega}
\newcommand{\vph}{\varphi}
\newcommand{\sig}{\sigma}
%%%%%%%%%%%%%%%%%%%%%%%%
%%%%%%%  C, R, Q, Z, N, Id  %%%%%
%%%%%%%%%%%%%%%%%%%%%%%%
%---- CORPS DES COMPLEXES
\newcommand{\CC}{\mbox{${\mathbb C}$}}
%---- CORPS DES REELS
\newcommand{\RR}{\mbox{${\mathbb R}$}}
%---- CORPS DES RATIONNELS
\newcommand{\QQ}{\mbox{${\mathbb Q}$}}
%---- GROUPE DES ENTIERS
\newcommand{\ZZ}{\mbox{${\mathbb Z}$}}
%---- NATURELS
\newcommand{\NN}{\mbox{${\mathbb N}$}}
%---- IDENTITE EN 12 PT
\newcommand{\1}{\mbox{\hspace{.0em}1\hspace{-.24em}I}}
\newcommand{\II}{\mbox{${\mathbb I}$}}
%%%%%%%%%%%%%%%%%%%%%%%%%%%%%%%%%
%%%%%%%%    DIVERS   %%%%%%%%%%%%
%%%%%%%%%%%%%%%%%%%%%%%%%%%%%%%%%
\newcommand{\prt}{\partial}
\newcommand{\und}[1]{\underline{#1}}
\newcommand{\wh}[1]{\widehat{#1}}
\newcommand{\wt}[1]{\widetilde{#1}}
\newcommand{\mb}[1]{\ \mbox{\ #1\ }\ }
\newcommand{\half}{\frac{1}{2}}
\newcommand{\noin}{\not\!\in}
\newcommand{\rhotimes}{\mbox{\raisebox{-1.2ex}{$\stackrel{\displaystyle\otimes}
{\mbox{\scriptsize{$\rho$}}}$}}}
\newcommand{\bin}[2]{{\left( {#1 \atop #2} \right)}}
\newcommand{\A}{{\cal A}}
\newcommand{\B}{{\cal B}}
\newcommand{\C}{{\cal C}}
\newcommand{\F}{{\cal F}}
\newcommand{\E}{{\cal E}}
\newcommand{\cP}{{\cal P}}
\newcommand{\R}{{\cal R}}
\newcommand{\T}{{\cal T}}
\newcommand{\W}{{\cal W}}
\newcommand{\cS}{{\cal S}}
\newcommand{\bS}{{\bf S}}
\newcommand{\cL}{{\cal L}}
\newcommand{\hlp}{{\RR}_+}
\newcommand{\hlm}{{\RR}_-}
\newcommand{\Hil}{{\cal H}}
\newcommand{\D}{{\cal D}}
\newcommand{\G}{{\cal G}}
\newcommand{\alg}{\C} 
\newcommand{\rep}{\F(\C)}
\newcommand{\trep}{\G_\beta(\C)}
\newcommand{\form}{\langle \, \cdot \, , \, \cdot \, \rangle }
\newcommand{\e}{{\rm e}}
\newcommand{\by}{{\bf y}}
\newcommand{\bp}{{\bf p}}
\newcommand{\LL}{\mbox{${\mathbb L}$}}
\newcommand{\Rp}{{R^+_{\, \, \, \, }}}
\newcommand{\Rm}{{R^-_{\, \, \, \, }}}
\newcommand{\Rpm}{{R^\pm_{\, \, \, \, }}}
\newcommand{\Tp}{{T^+_{\, \, \, \, }}}
\newcommand{\Tm}{{T^-_{\, \, \, \, }}}
\newcommand{\Tpm}{{T^\pm_{\, \, \, \, }}}
\newcommand{\baral}{\bar{\alpha}}
\newcommand{\barbt}{\bar{\beta}}
\newcommand{\supp}{{\rm supp}\, }
\newcommand{\EE}{\mbox{${\mathbb E}$}}
\newcommand{\JJ}{\mbox{${\mathbb J}$}}
\newcommand{\MM}{\mbox{${\mathbb M}$}}
\newcommand{\ct}{{\cal T}}

%%%%%%%% THEOREMES
\newtheorem{theo}{Theorem}[section]
\newtheorem{coro}[theo]{Corollary}
\newtheorem{prop}[theo]{Proposition}
\newtheorem{defi}[theo]{Definition}
\newtheorem{conj}[theo]{Conjecture}
\newtheorem{lem}[theo]{Lemma}
\newcommand{\prf}{\underline{\it Proof.}\ }
\newcommand{\finprf}{\null \hfill {\rule{5pt}{5pt}}\\[2.1ex]\indent}

%%%%%%%%%%%%%%%%%%%%%%%
\pagestyle{empty}
\rightline{May 2004}

\vfill

\begin{center}
{\Large\bf Finite Temperature Quantum Field Theory\\ with Impurities}
\\[2.1em]

\bigskip

{\large
M. Mintchev$^{a}$\footnote{mintchev@df.unipi.it} 
and P. Sorba$^{b}$\footnote{sorba@lapp.in2p3.fr}}\\

\null

\noindent 

{\it $^a$ INFN and Dipartimento di Fisica, Universit\'a di
      Pisa, Via Buonarroti 2, 56127 Pisa, Italy\\[2.1ex]
$^b$ LAPTH, 9, Chemin de Bellevue, BP 110, F-74941 Annecy-le-Vieux
      cedex, France}
\vfill

\end{center}

\begin{abstract} 
We apply the concept of reflection-transmission (RT) algebra, originally developed in the 
context of integrable systems in 1+1 space-time dimensions, 
to the study of finite temperature quantum field theory with impurities 
in higher dimensions. We consider a scalar field in $(s+1)+1$ space-time dimensions, 
interacting with impurities localized on $s$-dimensional hyperplanes, but 
without self-interaction. We discuss first the case $s=0$ 
and extend afterwards all results to $s>0$. 
Constructing the Gibbs state over an appropriate RT algebra, we derive the energy 
density at finite temperature and establish the correction 
to the Stefan-Boltzmann law generated by the 
impurity. The contribution of the impurity bound states is taken into account. 
The charge density profiles for various impurities are also investigated.  

\end{abstract}

\vfill
\rightline{IFUP-TH 21/2004}
\rightline{LAPTH-1045/04}
\rightline{\tt hep-th/0405264}
\newpage
\pagestyle{plain}
\setcounter{page}{1}

%%%%%%%%%%%%%%%%%%%%%%%%%%%%%%%%
\sect{Introduction}

Impurities (defects) exhibit a number of intriguing physical 
and mathematical features. Due to the rapid progress in building nanoscale 
quantum devises, they find nowadays interesting physical applications. 
These facts have motivated several investigations in the context of 
quantum mechanics \cite{A1} -- \cite{SCH} and quantum field theory (QFT) 
\cite{Delfino:1994nr} -- \cite{Langmann}. 

In some recent papers \cite{Mintchev:2002zd} -- \cite{Mintchev:2003kh} we have 
proposed an algebraic framework for dealing with defects in 1+1 space-time dimensions, 
introducing the so called {\it reflection-transmission} (RT) algebras. 
We have shown that these algebras 
represent a powerful approach to integrable systems with 
impurities, allowing to reconstruct the total scattering operator from the basic 
scattering data, namely the two-body bulk $S$-matrix and the reflection and transmission 
amplitudes of a single particle interacting with the impurity. 

In the present article we pursue further our analysis, extending the framework of 
RT algebras in two directions. We show that RT algebras are well adapted 
for treating impurity problems in higher space-time dimensions. 
Moreover, we generalize the framework to finite temperatures. Both 
these generalizations are essential for realistic 
applications to condensed matter physics. The most natural
way to perform the extension to finite temperature  
is to substitute the Fock vacuum, used in scattering 
theory, with a Gibbs grand canonical 
equilibrium state at given (inverse) temperature $\beta$. The Gibbs state is 
required to satisfy the Kubo-Martin-Schwinger (KMS) condition and 
defines a new (non-Fock) representation of the underlying RT algebra. 
This representation is the main tool for our analysis.  
We consider the case without self-interactions and focus essentially on 
two observables -- the energy-momentum tensor 
$\theta_{\mu \nu}$ and the conserved $U(1)$-current $j_\mu$. Computing 
the expectation value of $\theta_{00}$ in the Gibbs state, 
we derive the thermal energy density and 
establish the corrections to the Stefan-Boltzmann law generated by the 
defect. The study of $j_\mu$ reveals on the other hand that the impurity 
affects also the charge distribution in the Gibbs state, but does not 
induce a persistent current. 

We start the paper by performing the quantization of a scalar field in the background 
of a general point-like defect in 1+1 space-time dimensions. For this purpose we use 
an appropriate RT algebra $\alg$. In section 3 we construct 
the finite temperature representation of $\alg$ and derive the thermal real time 
correlation functions, obeying the KMS condition. In section 4 we compute the 
energy density and establish the corrections to the Stefan-Boltzmann law 
induced by the impurity. The charge and current densities are 
investigated in section 5. For illustration we describe there the charge density profiles 
corresponding to various impurities.  Section 6 is devoted to the impact of 
impurity bound states. In section 7 we generalize the 
results of the previous sections about point-like impurities in 1+1 space-time 
dimensions to $s$-dimensional hyperplane-defects in $(s+1)+1$ dimensions. 
We discuss in particular plane-defects in the physical 3+1 dimensions. 
The last section collects our conclusions and comments 
about the further research in the subject. 

\bigskip

\sect{General point-like defects} 

Employing the concept of RT algebra, we extend 
in this section some basic results \cite{A1}--\cite{SCH} on point-like impurities in 
quantum mechanics to QFT in 1+1 dimensions. We consider an impurity localized at $x=0$ and 
a hermitian scalar quantum field $\varphi(t,x)$ satisfying 
\be
[\prt_t^2 - \prt_x^2 + m^2] \varphi (t,x) = 0\, , \qquad x\not= 0 \, ,  
\label{eqm}
\ee
with the standard initial conditions:
\be
[\varphi (0,x_1)\, ,\, \varphi (0,x_2)] = 0\, , \qquad
[(\prt_t\varphi )(0,x_1)\, ,\, \varphi (0,x_2)] = -i\delta (x_1-x_2) \, .
\label{initial}
\ee 
It is worth stressing that the equation of motion (\ref{eqm}) is not imposed 
on the impurity. For this reason the solution of eqs. (\ref{eqm},\ref{initial}) is not 
unique, contrary to the case in which eq. (\ref{eqm}) holds for any $x\in \RR$. 
The physical problem of describing all admissible point-like impurities 
is equivalent to the mathematical problem of classifying the possible solutions 
of eqs. (\ref{eqm},\ref{initial}). In order to solve the latter, 
one has to analyse the operator $-\prt_x^2$, defined on the space 
$C_0^\infty (\RR \setminus \{0\})$ of smooth functions with compact support separated 
from the origin $x=0$. This operator is not self-adjoint, but its closure admits 
self-adjoint extensions.  It has been shown in \cite{A1,A2} that all of them are parametrized  by 
\begin{equation}  
\{(a,\, b,\, c,\, d;\, \varepsilon) \, :\, ad -bc = 1,\, 
{\overline \varepsilon} \varepsilon = 1,\,  
a,...,d \in \RR,\, \varepsilon \in \CC \} \, .  
\nonumber 
\end{equation}  
Since $\varphi$ is hermitian, in our case $\varepsilon \in \RR$ and therefore 
$\varepsilon = \pm 1$. The value $\varepsilon = -1$ can be absorbed in $(a,b,c,d)$ 
and we are thus left with 
\be 
\Gamma = \{\gamma = (a,\, b,\, c,\, d) \, :\, ad -bc = 1,\,  a,...,d \in \RR \} \, .  
\label{parameters}
\ee 
Each $\gamma \in \Gamma$ defines a {\it unique} 
solution of eqs. (\ref{eqm},\ref{initial}),  which obeys 
\be
\left(\begin{array}{cc} \varphi (t,+0) \\ \prt_x \varphi (t,+0)\end{array}\right) = 
\left(\begin{array}{cc} a & b\\ c&d\end{array}\right)
\left(\begin{array}{cc} \varphi (t,-0) \\ \prt_x \varphi (t,-0)\end{array}\right) 
\label{bc}
\ee 
{}for any $t\in \RR$. We observe that the 
impurity boundary conditions (\ref{bc}) can be implemented, coupling the 
field $\varphi$ to an external potential with support in $x=0$. The quadruple 
\be 
\gamma_\eta = 
(1,\, 0,\, 2\eta,\, 1) \, , \qquad \eta \in \RR \,  , 
\label{deltaimp}
\ee 
{}for instance, corresponds to the potential 
\be
V(x) = 2\eta \delta (x)\, ,   
\label{potential}
\ee
known \cite{A1} as $\delta$-impurity. The conventional free scalar field is 
obtained for $\eta = 0$. A potential incorporating all parameters 
(\ref{parameters}) has been recently proposed in \cite{Langmann}. 

A fundamental input in the quantization of $\varphi$ is the spectrum of the self-adjoint 
extension $\{-\prt_x^2,\, \gamma \}$. It is convenient to distinguish the following 
three subdomains of $\Gamma$:  
\bea 
\Gamma_0 = \{b<0,\, r_+\geq 0\}\cup  
\{b=0,\, r_0 \geq 0\}\cup\{b>0,\, r_- \geq 0\}\, , \qquad \\ 
\Gamma_1 = \{b<0,\, r_+<0\leq r_- \}\cup  
\{b=0,\, r_0 < 0\}\cup  \{b>0,\, r_-<0\leq r_+ \}\, , \quad \quad  \\ 
\Gamma_2 =  \{b<0,\, r_-<0\}\cup \{b>0,\, r_+<0\} \, , \qquad \qquad 
\eea 
where 
\be 
r_0 = \frac{c}{a+d}\, , \qquad 
r_\pm = \frac{a+d \pm \sqrt {(a-d)^2 + 4}}{2b} \, .  
\ee 
By construction 
\be 
\Gamma = \Gamma_0 \cup \Gamma_1 \cup \Gamma_2 \, , 
\label{decomposition}
\ee 
the index $n$ of $\Gamma_n$ indicating the number of bound states present in the 
spectrum of $\{-\prt_x^2,\, \gamma\in \Gamma_n \}$. We devote 
section 6 to the case $\gamma \in \Gamma_1 \cup \Gamma_2 $, focusing 
in the rest of the paper on the case without bound states .  

A complete orthonormal  system of scattering states for 
$\gamma \in \Gamma_0$ is given by 
\be
\psi_k^+(x) = \theta(-k)\left \{\theta(-x) T_-^+(k) \e^{ikx} +
\theta(x)\left [\e^{ikx} + R_+^+(-k)\e^{-ikx}\right ]\right \}\, , 
\label{basis1}
\ee
\be
\psi_k^-(x) = \theta(k)\left\{\theta(x) T_+^-(k) \e^{ikx} +
\theta(-x)\left [\e^{ikx} + R_-^-(-k)\e^{-ikx}\right ]\right \}\, , 
\label{s}
\ee
where $\theta$ denotes the standard Heaviside function and 
\bea
R_+^+(k) = \frac{bk^2 + i(a-d)k + c}{bk^2 + i(a+d)k - c} \, , \qquad 
T_+^-(k) = \frac{2ik}{bk^2 + i(a+d)k - c}\, , 
\label{coef1} \\
R_-^-(k) = \frac{bk^2 + i(a-d)k + c}{bk^2 - i(a+d)k - c} \, , \qquad 
T_-^+(k) = \frac{-2ik}{bk^2 - i(a+d)k - c}\, , 
\label{coef2}
\eea 
are the {\it reflection} and {\it transmission coefficients} from the impurity.  
It is easily verified that the {\it reflection} and {\it transmission matrices}, 
defined by 
\be 
\R(k) = \left(\begin{array}{cc} R_+^+(k) & 0\\0 & R_-^-(k)\end{array}\right)\, , 
\qquad 
\T(k) =\left(\begin{array}{cc} 0 & T_+^-(k)\\T_-^+(k) & 0\end{array}\right)\, , 
\label{rtmat}
\ee
satisfy hermitian analyticity 
\be 
\R(k)^\dagger = \R(-k)\,  , \qquad  \T(k)^\dagger = \T(k)\,  , 
\label{hanal}
\ee 
and unitarity 
\bea
\T(k) \T(k) + \R(k) \R(-k)  = \II \, , 
\label{unit1} \\
\T(k) \R(k) +  \R(k) \T(-k) = 0 \,  . 
\label{unit2}
\eea 

{}Following \cite{Mintchev:2003ue}, for $\gamma \in \Gamma_0$ 
we introduce the decomposition 
\be
\varphi (t,x) = \varphi_+ (t,x) + \varphi_- (t,x)  \,  ,
\label{ff1}
\ee
setting  
\be
\varphi_\pm (t,x) = \int_{-\infty}^{+\infty} \frac{dk}{2\pi \sqrt
{2\omega (k)}}
\left[a^{\ast \pm}(k) {\overline \psi}_k^{\, \pm} (x)\e^{i\omega (k)t} +
a_\pm (k) \psi_k^\pm (x)\e^{-i\omega (k)t}\right ] \,  ,
\label{ff2}
\ee 
where $\omega (k) = \sqrt {k^2 + m^2}$ and 
$\{a^{\ast \xi} (k),\, a_\xi (k)\, :\, \xi=\pm,\, k\in \RR \}$ 
generate the bosonic RT algebra $\alg$ with identity element $\bf 1$:  
\bea
&a_{\xi_1}(k_1)\, a_{\xi_2}(k_2) -  a_{\xi_2}(k_2)\, a_{\xi_1}(k_1) = 0\,  ,
\label{ccr1} \\
&a^{\ast \xi_1}(k_1)\, a^{\ast \xi_2}(k_2) - a^{\ast \xi_2}(k_2)\,
a^{\ast \xi_1}(k_1) = 0\,  ,
\label{ccr2} \\
&a_{\xi_1}(k_1)\, a^{\ast \xi_2}(k_2) - a^{\ast \xi_2}(k_2)\,
a_{\xi_1}(k_1) = \nonumber \\
&\left [\delta_{\xi_1}^{\xi_2} + \T_{\xi_1}^{\xi_2}(k_1)\right ] 2\pi
\delta(k_1-k_2)\, {\bf 1} +
\R_{\xi_1}^{\xi_2}(k_1) 2\pi \delta(k_1+k_2)\, {\bf 1}\,  .
\label{ccr3}
\eea 
The presence of the impurity is captured by the reflection and transmission 
matrices (\ref{rtmat}), which appear in the right-hand side of eq. (\ref{ccr3}) 
and by the constraints 
\bea
a_\xi(k) &=& \T_\xi^\eta (k) a_\eta (k) + \R_\xi^\eta (k) a_\eta (-k) \, ,
\label{c1} \\
a^{\ast \xi}(k) &=& a^{\ast \eta}(k) \T_\eta^\xi (k) +
a^{\ast \eta}(-k) \R_\eta^\xi (-k) \, ,
\label{c2}
\eea
imposed on the generators of $\alg$. Taking into account (\ref{c1},\ref{c2}), the fields 
$\varphi_\pm$ can be rewritten as 
\be
\varphi_\pm (t,x) = \theta(\pm x) \int_{-\infty}^{+\infty} \frac{dk}{2\pi \sqrt
{2\omega (k)}}
\left[a^{\ast \pm}(k) \e^{i\omega (k)t-ikx} +
a_\pm (k) \e^{-i\omega (k)t+ikx}\right ] \,  . 
\label{ff3}
\ee 
The fact that $\varphi_\pm$ satisfy the defect boundary conditions (\ref{bc}) is 
less explicit in this form, which is suitable however for the generalization to 
higher space-time dimensions described in section 7. 

All details about the structure defined by equations (\ref{ccr1})--(\ref{c2}) 
and their origin are given in \cite{Mintchev:2003ue}. The reflection and 
transmission matrices, appearing in the right hand side of (\ref{ccr3}), capture 
the interaction with the impurity. The constraints (\ref{c1}, \ref{c2}) are 
a consequence of unitarity (\ref{unit1}, \ref{unit2}) and of a peculiar 
(called reflection-transmission) authomorphism of the RT algebra $\alg $, 
established in \cite{Mintchev:2003ue}. 
The fields $\varphi_\pm$, defined by (\ref{ff3}), are elements of $\alg$. 
In order to obtain quantum fields acting in a Hilbert space, one must 
take a representation of $\alg$. In \cite{Mintchev:2003ue} 
we have constructed the Fock representation 
$\rep$, showing that the asymptotic in- and out-spaces of $\varphi$ coincide 
with $\rep$. The basic correlator of $\varphi$ in $\rep$ is the 
two-point vacuum expectation value. It is obtained 
employing (\ref{ccr3}) and the fact that $a_\xi (k)$ annihilates 
the vacuum state $\Omega \in \rep$. One gets 
\be
\langle  \varphi (t_1,x_1)\varphi (t_2,x_2)\rangle_\Omega  = 
\int_{-\infty}^{+\infty} \frac{dk}{4\pi\omega (k)} \e^{-i\omega (k)t_{12}}  
E(k;x_1,x_2;\gamma ) \, ,
\label{w20} 
\ee 
where 
\bea
E(k;x_1,x_2;\gamma) = 
\theta (x_1)\theta (-x_2) T_+^-(k)\e^{ikx_{12}} + 
\theta (-x_1)\theta (x_2) T_-^+(k)\e^{ikx_{12}} + 
\nonumber \\
\theta (x_1)\theta (x_2)\left [\e^{ikx_{12}} + R_+^+(k)\e^{ik{\widetilde x}_{12}}\right ]+
\theta (-x_1)\theta (-x_2)\left [\e^{ikx_{12}} + R_-^-(k)\e^{ik{\widetilde x}_{12}}\right] \, ,  
\label{ew20}
\eea
with $x_{12} = x_1-x_2$ and ${\widetilde x}_{12} = x_1+x_2$. As expected, for 
the free field $\gamma_0 = (1,0,0,1)$ the above expression gives 
$$ 
E(k;x_1,x_2;\gamma_0) = \e^{ikx_{12}} \, . 
$$ 

The symmetry content of our system can be deduced from 
(\ref{w20},\ref{ew20}). First of all 
(\ref{w20}) is invariant under time translations, which implies energy conservation. 
In spite of the relativistic dispersion relation $\omega (k)^2 = k^2 + m^2$, 
each non-trivial ($\gamma \not= \gamma_0$) impurity violates Lorentz and 
space translation invariance. The reflection $x\mapsto -x$ 
leaves invariant eqs. (\ref{eqm},\ref{initial}), but not always (\ref{bc}). 
In fact, (\ref{w20}) is invariant under spatial reflections if and only if 
\be 
a=d\, . 
\label{parity}
\ee
This condition selects the parity preserving impurities. 

Let us observe in conclusion that the Hamiltonian $H$, generating the time evolution 
\be 
\varphi(t,x) = \e^{itH} \varphi (0,x) \e^{-itH} \,  ,  \nonu 
\ee 
is given by the familiar quadratic expression 
\be 
H = \sum_{\xi = \pm} \int_{-\infty}^{+\infty} 
\frac{dk}{2\pi}\omega(k) a^{\ast\xi}(k) a_\xi (k) \, , 
\label{ham}
\ee  
which will be  adopted in the next section for the construction of 
the Gibbs state over the RT algebra $\alg$. 

\bigskip 

\sect{The Gibbs state over $\alg$} 

Our goal here is to construct a thermal representation $\trep$ of 
$\alg$. For this purpose we introduce $K\in \alg$ defined by 
\be
K = H - \mu N \, , \qquad \mu \in \RR \, , 
\label{K}
\end{equation}
where $H$ is the Hamiltonian (\ref{ham}), $\mu$ is the chemical potential and 
$N$ is the number operator 
\be
N = \sum_{\xi = \pm} \int_{-\infty}^{+\infty} 
\frac{dk}{2\pi} a^{\ast\xi}(k) a_\xi (k) \, . 
\label{num}
\ee 
We choose as a cyclic vector, 
determining $\trep$, the Gibbs grand canonical equilibrium state 
corresponding to $K$ at inverse temperature $\beta \equiv T^{-1}$. 
Differently from the  vacuum $\Omega \in \rep$, the Gibbs
state is not annihilated by 
$a_\xi(k)$ and represents an appropriate idealization of a thermal bath, 
keeping our system in equilibrium. The expectation value of a generic 
polynomial $\cP$ in this state is defined by 
\be
\langle \cP(a^{\ast \xi_i}(p_i), a_{\zeta_j}(q_j)) \rangle_\beta  = 
\frac{{\rm Tr}\left [\e^{-\beta K}\cP(a^{\ast \xi_i}(p_i), a_{\zeta_j}(q_j))\right ]}
{{\rm Tr}\, \e^{-\beta K}} \, . 
\label{generic} 
\end{equation}
Assuming the normalization condition
\be
\langle {\bf 1} \rangle_\beta   = 1  
\label{nc}
\end{equation} 
{}for the identity element ${\bf 1}\in \alg$, the non-vanishing two-point
correlation functions in the Gibbs state are given by
\bea
\langle a^{\ast \xi_1}(p_1)a_{\xi_2}(p_2)\rangle_\beta  =
\frac{\e^{-\beta [\omega(p_1)-\mu]}}{ 1-\e^{-\beta [\omega(p_1)-\mu]}} \cdot 
\qquad \qquad \nonumber \\
\left\{\left[\delta_{\xi_2}^{\xi_1} + \T_{\xi_2}^{\xi_1}(p_1)\right]2\pi \delta (p_1-p_2) 
+ \R_{\xi_2}^{\xi_1}(-p_1)2\pi \delta (p_1+p_2)  \right \}\,  ,
\label{be1}
\\ 
\langle a_{\xi_1}(p_1)a^{\ast \xi_2}(p_2)\rangle_\beta  =
\frac{1}{ 1-\e^{-\beta [\omega(p_1)-\mu]}} \cdot 
\qquad \qquad \nonumber \\
\left\{\left[\delta_{\xi_2}^{\xi_1} + \T_{\xi_2}^{\xi_1}(p_1)\right]2\pi \delta (p_1-p_2) 
+ \R_{\xi_2}^{\xi_1}(p_1)2\pi \delta (p_1+p_2)  \right \}\,  .
\label{be2}
\eea
Notice that the Bose-Einstein distribution appears as a factor in the right hand side of
(\ref{be1},\ref{be2}). 

It is instructive to sketch here a simple derivation of the correlators 
(\ref{be1},\ref{be2}), generalizing the argument in \cite{BR} for the 
case without impurities. Since (\ref{be1},\ref{be2}) are related by the
commutator (\ref{ccr3}), it is enough to concentrate on (\ref{be1}) for instance. 
Using the identity 
\be 
\e^{-\beta K} a^{\ast \xi}(k) = 
\e^{-\beta[\omega(k) - \mu]}\, a^{\ast \xi}(k)\, \e^{-\beta K} \, , 
\ee 
the cyclicity of the trace and the commutation relation (\ref{ccr3}), one finds
\bea
\langle a^{\ast \xi_1}(p_1)a_{\xi_2}(p_2)\rangle_\beta  = 
\frac{{\rm Tr}\left [a^{\ast \xi_1}(p_1)\e^{-\beta K}a_{\xi_2}(p_2)\right ]}
{{\rm Tr}\, \e^{-\beta K}}\, \e^{-\beta [\omega(p_1)-\mu]} = 
\qquad \nonumber \\
\frac{{\rm Tr}\left [\e^{-\beta K}a_{\xi_2}(p_2)a^{\ast \xi_1}(p_1)\right ]}
{{\rm Tr}\, \e^{-\beta K}} \e^{-\beta [\omega(p_1)-\mu]} = 
\e^{-\beta [\omega(p_1)-\mu]} 
\langle a^{\ast \xi_1}(p_1)a_{\xi_2}(p_2)\rangle_\beta  + 
\nonumber \\ 
\e^{-\beta [\omega(p_1)-\mu]} \left\{\left[\delta_{\xi_2}^{\xi_1} +
\T_{\xi_2}^{\xi_1}(p_1)\right]2\pi \delta (p_1-p_2)  + \R_{\xi_2}^{\xi_1}(-p_1)2\pi \delta
(p_1+p_2)  \right \},    
\label{derivation}
\eea 
which represents an equation for 
$\langle a^{\ast \xi_1}(p_1)a_{\xi_2}(p_2)\rangle_\beta $. 
The solution of this equation gives (\ref{be1}). In general, any correlator (\ref{generic}) 
can be evaluated by iteration, applying the identity 
\be 
\langle \prod_{i=1}^m a^{\ast \xi_i}(p_i) \prod_{j=1}^n a_{\zeta_j}(q_j)
\rangle_{\beta} = 
\delta_{mn}\, \sum_{k=1}^m \langle a^{\ast \xi_1}(p_1)a_{\zeta_k}(q_k)\rangle_{\beta} 
\, \langle \prod_{i=2}^m a^{\ast \xi_i}(p_i) \prod_{\stackrel{j=1}{j\not=k} }^n a_{\zeta_j}(q_j) 
\rangle_{\beta}\, ,  
\label{2a}
\ee
and the commutation relations (\ref{ccr1}--\ref{ccr3}). Like in the Fock representation 
$\rep$, all correlators in $\trep$ with different number of creation and annihilation 
operators vanish. 

We have thus shown that eqs. (\ref{nc}--\ref{be2},\ref{2a}) completely determine the
thermal representation $\trep$ of the RT algebra $\alg$. One can analyse at this point the 
KMS condition in $\trep$. This condition relates the two-point correlators 
(\ref{be1},\ref{be2}) according to 
\be
\langle \left [\alpha_s a_{\xi_1}(p_1)\right ]a^{\ast \xi_2}(p_2)\rangle_\beta  =
\langle a^{\ast \xi_2}(p_2)\left [\alpha_{s+i\beta} a_{\xi_1}(p_1)\right ]\rangle_\beta  \, , 
\label{KMS}
\end{equation} 
$\alpha_s$ being the automorphism on $\alg$ 
\be
\alpha_s \, a^{\ast \xi}(k) = a^{\ast \xi}(k) \e^{is[\omega(k)- \mu]} \, , \qquad
\alpha_s \, a_\xi (k) = a_\xi (k) \e^{-is[\omega(k)- \mu]}\, ,   
\end{equation} 
generated by $K$. The identity (\ref{KMS}) can be verified directly, using the explicit form of the 
correlators. Eq. (\ref{2a}) then extends the validity of the KMS condition to the whole $\trep$. 

The decomposition (\ref{ff3}) allows to determine 
all correlation functions of $\varphi$ in $\trep$. The two-point function 
reads  
\bea
\langle  \varphi (t_1,x_1)\varphi (t_2,x_2)\rangle_\beta   = 
\qquad \qquad \qquad \qquad \nonumber \\ 
\int_{-\infty}^{+\infty} \frac{dk}{4\pi\omega (k)} \left \{
\frac{\e^{-\beta [\omega(k)-\mu]+i\omega (k)t_{12}} + \e^{-i\omega (k)t_{12}}}
{ 1-\e^{-\beta [\omega(k)-\mu]}} \right \} E(k;x_1,x_2;\gamma )\,  .  
\label{tw20}
\eea 
and comparing with eq. (\ref{w20}), we see that the passage from 
zero to finite temperature is equivalent to the substitution 
\be 
\nonumber 
\e^{-i\omega (k)t_{12}} \longmapsto 
\frac{\e^{-\beta [\omega(p_1)-\mu]+i\omega (k)t_{12}} +
\e^{-i\omega (k)t_{12}}} { 1-\e^{-\beta [\omega(p_1)-\mu]}} 
\ee 
in the integrand of (\ref{w20}). We observe also that in the limit $\beta \to \infty$ 
one recovers from (\ref{tw20}) the Fock space correlator (\ref{w20}), provided that 
$\mu \leq m$. 

Summarizing, we investigated above a hermitian scalar field 
interacting with a general point-like impurity in 1+1 space-time dimensions. 
We derived in explicit form the real time finite temperature correlation 
functions, which satisfy the KMS condition. Using this result, we will analyse 
in what follows the physical properties of the system under consideration. For simplicity 
we will consider mostly the case $\mu=0$. Non-vanishing chemical potentials can be 
dealt with along the same lines. 

\bigskip 

\sect{Energy density and Stefan-Boltzmann law} 

In order to derive the energy density in the Gibbs state, we first express 
the Hamiltonian (\ref{ham}) in terms of $\varphi$. One has 
\be 
H= \int_{-\infty}^{-0} dx\, \theta_{00}(t,x) +  \int^{+\infty}_{+0} dx\, \theta_{00}(t,x) \, , 
\label{h1}
\ee 
where $\theta_{00}$ is the energy density operator 
\be 
\theta_{00}(t,x) = \frac{1}{2}
\left [:\prt_t\varphi \prt_t\varphi :(t,x) - :\varphi \prt_x^2\varphi :(t,x) 
+ m^2 :\varphi \varphi :(t,x)\right ] 
\label{edensityop}
\ee 
and $:\cdots :$ denote the normal product in the algebra $\alg$. 
Using the conditions (\ref{bc}) at $x=\pm 0$, one can check directly 
that $H$ given by (\ref{h1}) is time independent. In this respect, the position of 
the derivative $\prt_x^2$ in the second 
term of $\theta_{00}$ is essential for matching the boundary terms, 
arising at $x=\pm 0$ from the integration by parts. The expectation value 
\be 
\E (x, \beta; \gamma ) = \langle \theta_{00}(t,x) \rangle_\beta \, , \qquad x\not= 0 \, , 
\label{edensity} 
\ee
represents the {\it thermal} energy density we are looking for. $\E$ is 
$t$-independent because the Gibbs state is invariant  under time translations. 

Because of the normal product in (\ref{edensityop}), $\E$ can be expressed in 
terms of the correlator (\ref{be1}). One finds 
\be 
\E (x, \beta; \gamma ) = 
\varepsilon_{{}_{\rm S-B}} (\beta) + 
\varepsilon (x, \beta; \gamma ) 
\label{edensity1} 
\ee 
with 
\be 
\varepsilon_{{}_{\rm S-B}} (\beta) = 
\int_{-\infty}^{+\infty} \frac{dk}{2\pi}
\frac{\omega (k) \e^{-\beta \omega(k)}} 
{ 1-\e^{-\beta \omega(k)}} \, ,   
\label{SB}
\ee
and 
\be  
\varepsilon (x, \beta; \gamma ) = 
\int_{-\infty}^{+\infty} \frac{dk}{2\pi}
\frac{\omega (k) \e^{-\beta \omega(k)}} 
{ 1-\e^{-\beta \omega(k)}} \left [\theta(x) R_+^+(k) + 
\theta(-x) R_-^-(k) \right ] \e^{2 i k x} \, . 
\label{d}
\ee 
$\E$ depends explicitly only on the reflection coefficients, but one should keep in 
mind the unitarity constraints (\ref{unit1},\ref{unit2}), which involve the transmission 
coefficients as well. Taking into account  
\be 
|R_+^+ (k)| \leq 1 \, , \qquad |R_-^- (k)| \leq 1\, , \qquad \forall \, k \in \RR \, , 
\label{est0}
\ee 
one easily derives from (\ref{d}) the estimate 
\be 
|\varepsilon (x, \beta; \gamma )| \leq \varepsilon_{{}_{\rm S-B}} (\beta) \, ,   
\label{est1}
\ee 
which implies 
\be 
\E (x, \beta; \gamma ) \geq 0 \, . 
\label{est2}
\ee 

The term (\ref{SB}) is present even without impurity and is actually the familiar 
Stefan-Boltzmann (S-B) contribution. One has for instance 
\be 
\varepsilon_{{}_{\rm S-B}} (\beta)|_{{}_{m=0}} = \frac{\pi}{6 \beta^2} \, , 
\label{SB1}
\ee 
which is the thermal energy density for a massless 
hermitian scalar field in 1+1 space-time dimensions. 
The term (\ref{d}) is of special interest because it  
describes the correction to the S-B law due to the defect. 
For getting a more precise idea about this correction, let us 
derive $\varepsilon$ for the 
$\delta$-impurity $\gamma_\eta^+ \equiv (1,0,2\eta >0,1)$. 
Taking for simplicity $m=0$, one has 
\be 
\varepsilon (x, \beta; \gamma_\eta^+ ) = -\frac{i\eta}{2\pi} 
\int_{-\infty}^{+\infty}d k 
\frac{|k| \e^{-\beta |k|}} { 1-\e^{-\beta |k|}} 
\frac{\e^{2 i k |x|}}{k+i\eta} \, ,  
\label{eta1}
\ee 
or equivalently, using the Feynman parameter $\alpha $ for representing 
the denominator $k+i\eta$, 
\be 
\varepsilon (x, \beta; \gamma_\eta^+) = 
-\frac{i\eta}{2\pi} 
\int_{-\infty}^{+\infty}d k \int_0^{+\infty}d \alpha 
\frac{|k| \e^{-\eta \alpha -\beta |k|}} { 1-\e^{-\beta |k|}} 
\e^{i k(2 |x|+\alpha)} \, .  
\label{eta2}
\ee
The exponential factor $\e^{-\eta \alpha -\beta |k|}$ in the integrand 
of (\ref{eta2}) allows to exchange the 
integrals in $k$ and $\alpha$. Integrating first over $k$ and then over $\alpha$ 
one gets \cite{PBM} 
\bigskip 
\bea 
\varepsilon (x, \beta; \gamma_\eta^+) = 
\frac{\eta}{2}\int_0^{+\infty}d \alpha \e^{-\eta \alpha} 
\left \{\frac{\pi}{\beta^2 \sinh^2 [\pi \beta^{-1}(2|x|+\alpha)]} 
- \frac{1}{\pi (2|x|+\alpha)^2} \right \} = 
\nonumber \\ 
\frac{2 \pi \eta \e^{-4 \pi \frac{|x|}{\beta}}}{\beta(2\pi + \beta \eta) }\, \,  
{}_2{\rm F}_1\left [2, 1+\frac{\beta \eta}{2 \pi};  2 +\frac{\beta \eta}{2 \pi}; 
\e^{-4 \pi \frac{|x|}{\beta}} \right ] 
- \frac{\eta}{4\pi |x|} - \frac{\eta^2 \e^{2|x|\eta}}{2 \pi}\, {\rm Ei}(-2|x|\eta), \quad 
\label{eta3} 
\eea
${}_2{\rm F}_1$ and ${\rm Ei}$ being the hypergeometric and 
exponential-integral functions respectively. The study of (\ref{eta3}) 
shows that the correction to the S-B law generated by the $\delta$-impurity 
is relevant close to the impurity. Moreover, $\varepsilon$ 
vanishes in both limits $\beta \to \infty$ and $|x| \to \infty$, as illustrated 
in Fig.1 and Fig.2 respectively.  
\begin{figure}[h]
\begin{center}
\includegraphics[width=6cm]{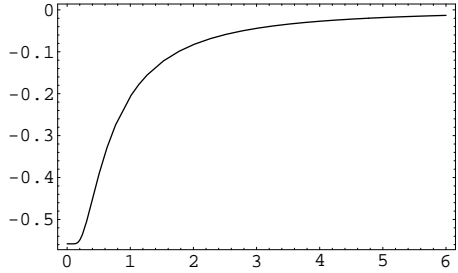}
\end{center}
\caption{Plot of $\varepsilon (0.1,\beta;\gamma_{\eta=1})$.} 
\label{f1}
\end{figure} 
\begin{figure}[h]
\begin{center}
\includegraphics[width=6cm]{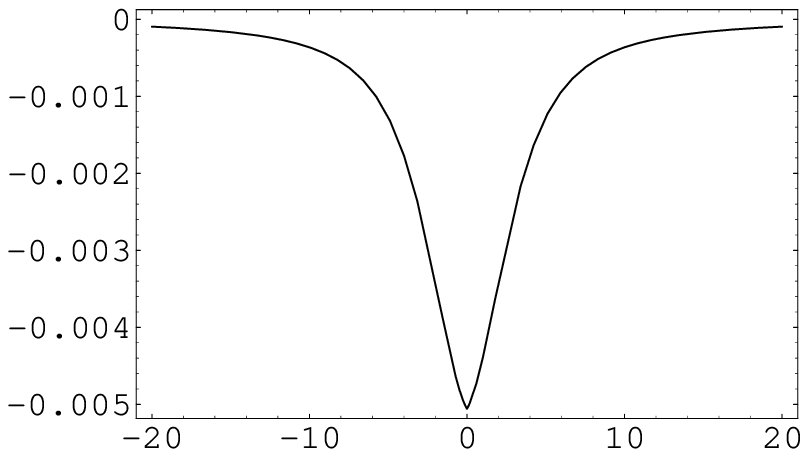}
\end{center}
\caption{Plot of $\varepsilon(x,10;\gamma_{\eta=1})$.}
\label{f2}
\end{figure} 

Combining (\ref{SB1}) with (\ref{eta3}), one gets 
\bea 
\E (x, \beta; \gamma_\eta^+ ) = 
\frac{\pi}{6 \beta^2} + 
\qquad \qquad  \qquad  \qquad  \qquad \nonumber \\ 
\frac{2 \pi \eta \e^{-4 \pi \frac{|x|}{\beta}}}{\beta(2\pi + \beta \eta) }\, \,  
{}_2{\rm F}_1\left [2, 1+\frac{\beta \eta}{2 \pi};  2 +\frac{\beta \eta}{2 \pi}; 
\e^{-4 \pi \frac{|x|}{\beta}} \right ] 
- \frac{\eta}{4\pi |x|} - \frac{\eta^2 \e^{2|x|\eta}}{2 \pi}\, {\rm Ei}(-2|x|\eta)\,  . \quad 
\label{eta4} 
\eea
By construction $\E$ collects only the thermal energy contributions and 
thus vanishes in the limit $T\to 0$. Technically, this feature is a consequence of 
the normal product in (\ref{edensityop}). Concerning the {\it total} energy density, one has 
\be 
\E_{\rm tot} (x, \beta; \gamma ) = \E (x, \beta ;\gamma ) + \E_{\rm C} (x; \gamma ) \, , 
\label{tot}
\ee 
where $\E_{\rm C}$ is the Casimir energy density at $T=0$, namely  
\be  
\E_{\rm C} (x; \gamma ) =  
\int_{-\infty}^{+\infty} \frac{dk}{4\pi}\, \omega(k) \left [\theta(x) R_+^+(k) + 
\theta(-x) R_-^-(k) \right ] \e^{2 i k x}   
\label{C}
\ee 
Eqs. (\ref{tot},\ref{C}) can be derived from (\ref{edensityop}), adopting 
the point splitting regularization instead of the normal product and 
subtracting the vacuum energy density of the free scalar field without 
impurity. For example, the Casimir energy density of the massless 
scalar field with the $\gamma_\eta^+$-impurity equals  
\be 
\E_{\rm C} (x; \gamma_\eta^+) = \frac{\eta}{4\pi |x|} +
\frac{\eta^2 \e^{2|x|\eta}}{2 \pi}\, {\rm Ei}(-2|x|\eta)\,  ,  
\label{etaC}
\ee 
which leads to the following total energy density
\be 
\E_{\rm tot} (x, \beta; \gamma_\eta^+) = 
\frac{\pi}{6 \beta^2} + 
\frac{2 \pi \eta \e^{-4 \pi \frac{|x|}{\beta}}}{\beta(2\pi + \beta \eta) }\, \,  
{}_2{\rm F}_1\left [2, 1+\frac{\beta \eta}{2 \pi};  2 +\frac{\beta \eta}{2 \pi}; 
\e^{-4 \pi \frac{|x|}{\beta}} \right ] \, . 
\ee

{}Finally, it is instructive to compare the total energy density with impurity to 
the energy density of the {\it free} scalar field $\gamma_0 = (1,0,0,1)$. 
Defining the pure impurity contribution by 
\be 
\E_{\rm imp} (x,\beta; \gamma  ) \equiv \E_{\rm tot} (x, \beta; \gamma  ) - 
\E_{\rm tot} (x, \beta; \gamma_0  )\, ,   
\label{pure} 
\ee
one gets 
\be 
\E_{\rm imp} (x,\beta; \gamma  ) = 
\int_{-\infty}^{+\infty} \frac{dk}{4\pi} \omega (k)
\frac{1+ \e^{-\beta \omega(k)}} 
{ 1-\e^{-\beta \omega(k)}} \left [\theta(x) R_+^+(k) + 
\theta(-x) R_-^-(k) \right ] \e^{2 i k x} \, .  
\label{genpure} 
\ee 
For the $\delta$-impurity one finds   
\bea 
\E_{\rm imp} (x,\beta; \gamma_\eta^+) =  
\frac{2 \pi \eta \e^{-4 \pi \frac{|x|}{\beta}}}{\beta(2\pi + \beta \eta) }\, \,  
{}_2{\rm F}_1\left [2, 1+\frac{\beta \eta}{2 \pi};  2 +\frac{\beta \eta}{2 \pi}; 
\e^{-4 \pi \frac{|x|}{\beta}} \right ] \, ,   
\label{etapure}
\eea 
which concludes our investigation of the energy density. 

\bigskip 

\sect{Charge and current densities} 

In this section we consider a complex scalar field 
\be 
\varphi (t,x) = \frac{1}{\sqrt 2} \left [\varphi_1 (t,x) + i \varphi_2 (t,x) \right ] \, , 
\ee
where both $\varphi_1$ and $\varphi_2$ are Hermitian 
scalar fields satisfying (\ref{eqm},\ref{initial}) and 
(\ref{bc}) with the same $\gamma \in \Gamma_0$. 
Our goal will be to derive the charge and current densities of the $U(1)$ conserved current 
\be 
j_\mu (t,x) = -i \left [:(\prt_\mu \varphi^\ast )\varphi :(t,x) - 
:\varphi^\ast (\prt_\mu \varphi ) :(t,x) \right ] \, . 
\label{current} 
\ee 
in the Gibbs state  constructed in sect. 3. Using the two-point function 
(\ref{tw20}), one obtains 
\bea 
\varrho(x,\beta;\gamma ) = \langle j_0(t,x) \rangle_\beta = 
\qquad \qquad \qquad \qquad \nonumber \\ 
\int_{-\infty}^{+\infty} \frac{dk}{\pi}
\frac{\e^{-\beta \omega(k)}} 
{ 1-\e^{-\beta \omega(k)}} \left \{1+\left [\theta(x) R_+^+(k) + 
\theta(-x) R_-^-(k) \right ] \e^{2 i k x}\right \} \, ,  
\label{cd}
\eea
\be 
\langle j_1(t,x) \rangle_\beta = 0\, . 
\label{currd}
\ee  
The coexistence of non-trivial charge density with 
vanishing current density is not 
surprising because the impurity breaks down Lorentz symmetry.  
Like for the energy density $\cal E$, one easily shows that 
\be 
\varrho(x,\beta;\gamma ) \geq 0 \, . 
\ee 

The temperature dependence of $\varrho$ is simple: 
$\varrho$ vanishes in the limit $T\to 0$ and diverges for $T\to \infty$. 
Concerning the space dependence, 
it is instructive to compare the $x$-distribution of the charge at a given 
temperature for different defects. Besides the already
familiar $\delta$-impurity, we will consider 
\be 
\gamma_\xi^+ = 
(1,\, 2\xi,\, 0,\, 1) \, , \qquad \xi > 0 \,  , 
\label{primeimp}
\ee 
and
\be 
\gamma_\zeta^+ = 
(2,\, 0,\, \zeta/2,\, 1/2)  \, , \qquad \zeta > 0 \,  . 
\label{paritybr}
\ee 
Inserting (\ref{deltaimp},\ref{primeimp},\ref{paritybr}) in (\ref{cd}) one finds 
\be 
\varrho(x,\beta;\gamma_\eta^+) = \int_0^{+\infty} \frac{dk}{\pi}\, 
\frac{2}{\e^{\beta \omega(k)}-1} \left \{1+
\frac{[\eta k \sin(2k|x|) -\eta^2 \cos(2k|x|)]}{k^2
+ \eta^2}\right \} \, ,   
\label{cddelta}
\ee 
\be 
\varrho(x,\beta;\gamma_\xi^+) = \int_0^{+\infty} \frac{dk}{\pi} \, 
\frac{2}{\e^{\beta \omega(k)}-1} \left \{1 + 
\frac{[\xi k \sin(2k|x|) +\xi^2 k^2 \cos(2k|x|)]}{\xi^2 k^2
+ 1}\right \} \, ,   
\label{cdprime}
\ee
and 
\bea
\varrho(x,\beta;\gamma_\zeta^+) = \int_0^{+\infty} \frac{dk}{\pi} \, 
\frac{2}{\e^{\beta \omega(k)}-1}
%\qquad \quad \nonumber \\
\Biggl \{1+ 
\theta(x) \frac{[2\zeta k \sin(2kx) +(k^2\zeta^2+15) \cos(2kx)]}{k^2\zeta^2 + 25} +
\nonumber \\ 
\theta(-x) \frac{[-8\zeta k \sin(2kx) +(k^2\zeta^2-15) \cos(2kx)]}{k^2\zeta^2 + 25} 
\Biggr \} \, ,   \qquad \qquad \qquad 
\label{cdparity}
\eea 
respectively. The densities (\ref{cddelta},\ref{cdprime}) 
are even functions in $x$ in agreement with the fact that 
the impurities $\gamma_\eta^+$ and $\gamma_\xi^+$ are parity preserving. 
The defect $\gamma_\zeta^+$ instead violates parity and gives rise to 
a distribution which is asymmetric with respect to the origin. 
\begin{figure}[h]
\begin{center}
\includegraphics[width=6cm]{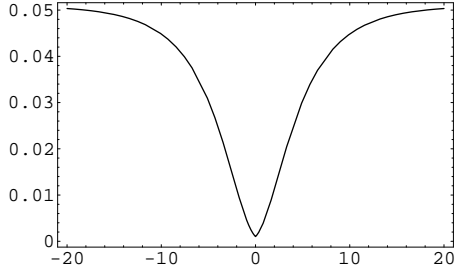}
\end{center}
\caption{Plot of $\varrho (x,10;\gamma_{\eta=1})$ for $m=0.1$.} 
\label{f3}
\end{figure}
\begin{figure}[h]
\begin{center}
\includegraphics[width=6cm]{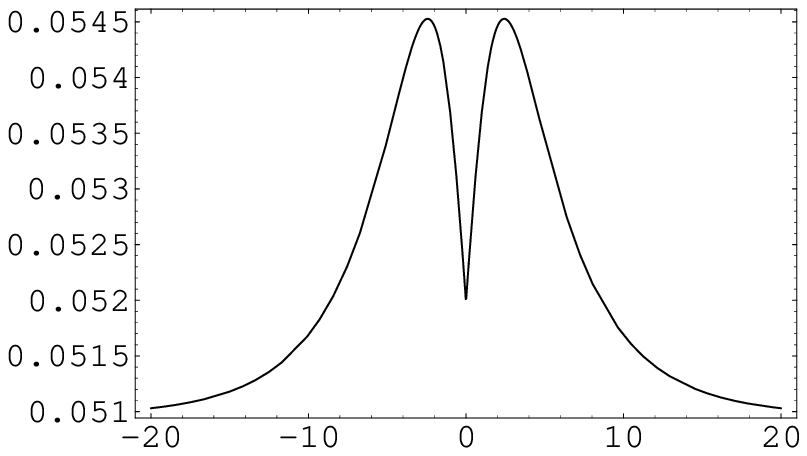}
\end{center}
\caption{Plot of $\varrho (x,10;\gamma_{\xi=1})$ for $m=0.1$.} 
\label{f4}
\end{figure} 
\begin{figure}[h]
\begin{center}
\includegraphics[width=6cm]{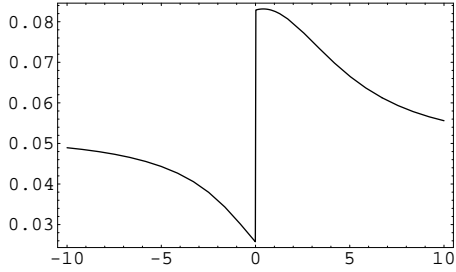}
\end{center}
\caption{Plot of $\varrho (x,10;\gamma_{\zeta=10})$ for $m=0.1$.} 
\label{f5}
\end{figure}  

To integrate in (\ref{cddelta}--\ref{cdparity}) over $k$ analytically   
is a hard job, but the numerical study gives a precise idea about 
the charge distribution. The Figures 3--5 display the shape of $\varrho$ as a 
function of $x$.  We see that the profile of $\varrho$ strongly 
depends on the type of impurity. For 
the $\delta$-defect the charge density (Fig.3) is minimal at the 
impurity. For $\gamma_\xi$ the distribution $\varrho$ has (Fig.4) two maxima 
localized away from the impurity. Finally, for $\gamma_\zeta$ one finds (Fig.5) that 
$\varrho$ is discontinuous in $x=0$  and reaches a minimum (maximum) when 
$x\to 0$ from the  left (right).  

The possibility to design the charge density profile by choosing appropriate 
$\gamma \in \Gamma$ looks very attractive for potential applications. 

\bigskip 

\sect{The contribution of impurity bound states} 

This section collects our results about the case 
$\gamma \in \Gamma_1\cup \Gamma_2$. 
The spectrum of $\{-\prt_x^2,\, \gamma \in \Gamma_1\}$ 
involves one bound state. 
A complete orthonormal system of eigenstates 
is given by (\ref{s}), supplemented by 
\be
 \chi (x) = \left\{\begin{array}{cc}
\sqrt{|r_+|}\, \e^{|x|r_+} \, ,
& \quad \mbox{$b<0$}\, ,\\[1ex]
\sqrt{|r_0|}\, \e^{|x|r_0}\, ,
& \quad \mbox{$b=0$}\, ,\\[1ex]
\sqrt{|r_-|}\, \e^{|x|r_-}\, ,
& \quad \mbox{$b>0$}\, . \\[1ex]
\end{array} \right.
\label{b1}
\end{equation} 
When $\gamma \in \Gamma_2$ one has two bound states, namely 
\be 
\chi_\pm(x) = \sqrt{|r_\pm|}\, \e^{|x|r_\pm} \,  .  
\label{b2}
\ee 
The states (\ref{b1},\ref{b2}) give rise to new quantum degrees of freedom 
for $\varphi$, which are represented by the last term in the decomposition 
\be
\varphi (t,x) = \varphi_+ (t,x) + \varphi_- (t,x) + \varphi_{\rm b} (t,x) \,  , 
\label{f1b}
\ee
where $\varphi_\pm$ are given by (\ref{ff3}). 
We emphasize that the contribution $\varphi_{\rm b}$ is uniquely fixed by 
the initial condition (\ref{initial}). If $\gamma \in \Gamma_1$ for instance, 
we must introduce the creation and annihilation operators $\{b^\ast\, ,\, b\}$,  
which commute with $\alg$ and satisfy
\be
[b\, ,\, b] = [b^\ast\, ,\, b^\ast] = 0\, , \qquad [b\, ,\, b^\ast ] = 1 \, . 
\label{osc}
\ee
One has in this case 
\be
\varphi_{\rm b} (t,x) = \frac{1}{\sqrt {2\omega}}
\left (b^\ast\, \e^{it\omega } +
b\, \e^{-it\omega }\right ) \chi  (x)\, ,
\label{f3b}
\ee
with 
\be
 \omega = \left\{\begin{array}{cc}
\sqrt{m^2-r_+^2} \, ,
& \quad \mbox{$b<0$}\, ,\\[1ex]
\sqrt{m^2-r_0^2}\, ,
& \quad \mbox{$b=0$}\, ,\\[1ex]
\sqrt{m^2-r_-^2}\, ,
& \quad \mbox{$b>0$}\, . \\[1ex]
\end{array} \right.
\label{eb1}
\end{equation} 
The two-point correlator reads 
\be 
w (t_{12}, x_1, x_2)\arrowvert_{{}_{\gamma \in \Gamma_1}} = 
w (t_{12}, x_1, x_2)\arrowvert_{{}_{\gamma \in \Gamma_0}} + 
\frac{1}{2\omega }\e^{-it_{12}\omega } \chi(x_1)  \chi(x_2) \, . 
\label{w21}
\ee 
Analogously, in the case $\gamma \in \Gamma_2$ one needs two 
oscillators $\{b^{\ast \pm}\, ,\, b_\pm\}$, both satisfying (\ref{osc}) 
and commuting with each other and with $\alg$. Now 
\be
\varphi_{\rm b} (t,x) = \sum_{\sigma = \pm} \frac{1}{\sqrt {2\omega_\sigma}}
\left (b^{\ast \sigma}\, \e^{it\omega_\sigma} +
b_\sigma \, \e^{-it\omega_\sigma }\right ) \chi_\sigma (x)\, , \qquad 
\omega_\pm = \sqrt{m^2-r_\pm ^2} \, , 
\label{ff4}
\ee
and 
\be 
w (t_{12}, x_1, x_2)\arrowvert_{{}_{\gamma \in \Gamma_2}} = 
w (t_{12}, x_1, x_2)\arrowvert_{{}_{\gamma \in \Gamma_0}} + 
\sum_{\sigma = \pm} \frac{1}{2\omega_\sigma }\e^{-it_{12}\omega_\sigma } 
\chi_\sigma(x_1)  \chi_\sigma(x_2)
\, .  
\label{w22}
\ee 
It follows from (\ref{eb1}) that in order to avoid imaginary energies and 
the associated quantum field instabilities, we must restrict further 
$\Gamma_1\cup\Gamma_2$ to 
\be 
{\widetilde \Gamma}_1 \cup {\widetilde \Gamma}_2 =  
\{\gamma \in \Gamma_1\cup\Gamma_2\, :\,  |r_0|\leq m,\, |r_\pm|\leq m\}\, . 
\ee
These restrictions are a consequence of the relativistic dispersion relation for 
$\varphi$ and are not present in the quantum mechanical context \cite{A1}--\cite{SCH}. 
Notice in particular that there are no defect bound state contributions for $m=0$, 
because both ${\widetilde\Gamma}_1$ and ${\widetilde \Gamma}_2$ are empty 
in this case. 

In the construction of the Gibbs state for 
$\gamma \in {\widetilde \Gamma}_1 \cup {\widetilde \Gamma}_2$ one must 
take into account the defect bound state degrees of freedom 
described by the oscillator algebras $\{b,\, b^\ast\}$ and $\{b_\pm,\, b_\pm^\ast\}$. 
The Hamiltonian is 
\be 
H = \sum_{\xi = \pm} \int_{-\infty}^{+\infty} 
\frac{dk}{2\pi}\omega(k) a^{\ast\xi}(k) a_\xi (k) + h_{\rm b} \, , 
\label{h0} 
\ee 
where 
\be 
h_{\rm b} = \left\{\begin{array}{cc} 
\omega b^\ast b\, ,
& \quad \mbox{$\gamma \in {\widetilde \Gamma}_1$}\, ,\\[1ex]
\sum_{\sigma = \pm} \omega_\sigma b^{\ast \sigma} b_\sigma\, ,
& \quad \mbox{$\gamma \in {\widetilde \Gamma}_2$}\, . \\[1ex]
\end{array} \right.
\label{hamb}
\end{equation} 
{}For the two-point functions at temperature $\beta$ and chemical potential 
$\mu$, one obtains 
\be 
w_\beta  (t_{12}, x_1, x_2)\arrowvert_{{}_{\gamma \in {\widetilde \Gamma}_1}} = 
w_\beta  (t_{12}, x_1, x_2)\arrowvert_{{}_{\gamma \in \Gamma_0}} + 
\frac{\e^{-\beta (\omega-\mu)+i\omega t_{12}} + \e^{-i\omega t_{12}}}
{ 2\omega [1-\e^{-\beta (\omega-\mu)}]} \chi(x_1)  \chi(x_2) \, ,  
\label{tw21}
\ee
and 
\bea 
w_\beta  (t_{12}, x_1, x_2)\arrowvert_{{}_{\gamma \in {\widetilde \Gamma}_2}} = 
w_\beta  (t_{12}, x_1, x_2)\arrowvert_{{}_{\gamma \in \Gamma_0}} + 
\nonumber \\ 
\sum_{\sigma = \pm}
\frac{\e^{-\beta (\omega_\sigma-\mu)+i\omega_\sigma t_{12}} + 
\e^{-i\omega_\sigma t_{12}}}
{ 2\omega_\sigma [1-\e^{-\beta (\omega_\sigma-\mu)}]} 
\chi_\sigma(x_1)  \chi_\sigma (x_2) \, . 
\label{tw22}
\eea
At this point we are ready to derive the impurity bound states correction to the 
S-B law. Instead of (\ref{edensity1}) one has 
\be 
\E (x, \beta; \gamma ) = 
\varepsilon_{{}_{\rm S-B}} (\beta) + 
\varepsilon (x, \beta; \gamma )  + \varepsilon_{{}_{\rm b}} (x, \beta; \gamma ) \, , 
\label{fdensity} 
\ee 
where 
\be 
\varepsilon_{{}_{\rm b}} (x, \beta; \gamma ) = \left\{\begin{array}{cc} 
\frac{\e^{-\beta \omega}} 
{ 1-\e^{-\beta \omega}}\, \omega\, \chi(x)^2\, ,
& \quad \mbox{$\gamma \in {\widetilde \Gamma}_1$}\, ,\\[1ex]
\sum_{\sigma = \pm}\frac{\e^{-\beta \omega_\sigma}} 
{ 1-\e^{-\beta \omega_\sigma}}\, \omega_\sigma\, \chi_\sigma (x)^2\, ,
& \quad \mbox{$\gamma \in {\widetilde \Gamma}_2$}\,  \\[1ex]
\end{array} \right.
\label{b}
\end{equation} 
is the bound state contribution. In view of (\ref{b1},\ref{b2}) this contribution decays 
exponentially with the distance from the impurity. The same property has the bound state 
correction to the charge density $\varrho$. 

\bigskip 

\sect{Defects in higher space-time dimensions} 

Our goal below is to show that all the results obtained in the previous sections can be
actually  generalized to higher  space-time dimensions. For this purpose we consider 
a $(s+1)+1$ dimensional Minkowski space, adopting the coordinates 
$(t,\, x,\, y )$ where $t, x\in \RR$ and 
$y \in \RR^s$. The impurity will be localized on a $s$-dimensional hyperplane 
in $x=0$. We impose on the quantum field $\varphi (t,x,y)$ the equation of motion 
\be
[\prt_t^2 - \prt_x^2 - \Delta_y + m^2] \varphi (t,x,y) = 0\, , \qquad x\not= 0 \, ,  
\label{heqm}
\ee 
the impurity boundary condition 
\be
\left(\begin{array}{cc} \varphi (t,+0,y ) \\ \prt_x \varphi (t,+0,y )\end{array}\right) = 
\left(\begin{array}{cc} a & b\\ c&d\end{array}\right)
\left(\begin{array}{cc} \varphi (t,-0,y) \\ \prt_x \varphi (t,-0,y )\end{array}\right)\, , 
\qquad \forall\, \,  t,\, y \in \RR  
\label{hbc}
\ee 
and the initial conditions   
\bea
\left [\varphi (0,x_1,y_1)\, ,\, \varphi (0,x_2,y_2)\right ] = 0\, , \qquad \qquad \qquad \\
\left [(\prt_t\varphi )(0,x_1,y_1)\, ,\, \varphi (0,x_2,y_2)\right ] = -i\delta (x_1-x_2) 
\delta(y_1-y_2)\, .
\label{hinitial}
\eea 
Eqs.(\ref{heqm}-\ref{hinitial}) have a unique solution. For $\gamma \in \Gamma_0$ one has 
\be
\varphi (t,x,y) = \varphi_+ (t,x,y) + \varphi_- (t,x,y)  \,  ,
\label{hf1}
\ee
\bea
\varphi_\pm (t,x,y) = \theta(\pm x) \int_{-\infty}^{+\infty} 
\frac{dk\, d^sp}{(2\pi)^{s+1}}\frac{1}{\sqrt{2\omega (k,p)}}\cdot 
\qquad \nonumber \\
\left[a^{\ast \pm}(k,p) \e^{i\omega (k,p)t-ikx-ip y} +
a_\pm (k,p) \e^{-i\omega (k,p)t+ikx+ip y}\right ] \,  . 
\label{hf3}
\eea 
Here $\omega (k,p) = \sqrt{k^2 +p^2 + m^2}$ and 
$\{a^{\ast \xi} (k,p),\, a_\xi (k,p)\, :\, \xi=\pm,\, k\in \RR,\, p\in\RR^s \}$ 
satisfy the following RT algebra commutation relations   
\bea
&a_{\xi_1}(k_1,p_1)\, a_{\xi_2}(k_2,p_2) -  
a_{\xi_2}(k_2,p_2)\, a_{\xi_1}(k_1,p_1) = 0\,  ,
\label{hccr1} \\
&a^{\ast \xi_1}(k_1,p_1)\, a^{\ast \xi_2}(k_2,p_2) - 
a^{\ast \xi_2}(k_2,p_2)\,
a^{\ast \xi_1}(k_1,p_1) = 0\,  ,
\label{hccr2} \\
&a_{\xi_1}(k_1,p_1)\, a^{\ast \xi_2}(k_2,p_2) - 
a^{\ast \xi_2}(k_2,p_2)\, a_{\xi_1}(k_1,p_1) = \nonumber \\
&\left\{ \left [\delta_{\xi_1}^{\xi_2} + \T_{\xi_1}^{\xi_2}(k_1)\right ] 
\delta(k_1-k_2) +
\R_{\xi_1}^{\xi_2}(k_1) \delta(k_1+k_2)\right \} (2\pi)^{s+1}\delta(p_1-p_2)\, {\bf 1}\,  
\label{hccr3}
\eea 
and the constraints 
\bea
a_\xi(k,p ) &=& \T_\xi^\eta (k) a_\eta (k,p) + \R_\xi^\eta (k) a_\eta (-k,p) \, ,
\label{hc1} \\
a^{\ast \xi}(k,p) &=& a^{\ast \eta}(k,p) \T_\eta^\xi (k) +
a^{\ast \eta}(-k,p) \R_\eta^\xi (-k) \, . 
\label{hc2}
\eea
The construction of the Gibbs state for this algebra follows that of section 3 and 
gives 
\bea
\langle a^{\ast \xi_1}(k_1,p_1)a_{\xi_2}(k_2,p_2)\rangle_\beta  =
\frac{\e^{-\beta [\omega(k_1,p_1)-\mu]}}{ 1-\e^{-\beta [\omega(k_1,p_1)-\mu]}} \cdot 
\qquad \qquad \nonumber \\
\left\{\left[\delta_{\xi_2}^{\xi_1} + \T_{\xi_2}^{\xi_1}(k_1)\right] \delta (k_1-k_2) 
+ \R_{\xi_2}^{\xi_1}(-k_1) \delta (k_1+k_2)  \right \}(2\pi )^{s+1}\delta(p_1-p_2)\,  ,
\label{hbe1}
\\ 
\langle a_{\xi_1}(k_1,p_1)a^{\ast \xi_2}(k_2,p_2)\rangle_\beta  =
\frac{1}{ 1-\e^{-\beta [\omega(k_1,p_1)-\mu]}} \cdot 
\qquad \qquad \nonumber \\
\left\{\left[\delta_{\xi_2}^{\xi_1} + \T_{\xi_2}^{\xi_1}(k_1)\right]\delta (k_1-k_2) 
+ \R_{\xi_2}^{\xi_1}(k_1)\delta (k_1+k_2)  \right \}(2\pi )^{s+1}\delta(p_1-p_2)\,  . 
\label{hbe2}
\eea 
The expectation values (\ref{hbe1},\ref{hbe2}) determine the finite temperature two-point
function 
\bea
\langle  \varphi (t_1,x_1,y_1)\varphi (t_2,x_2,y_2)\rangle_\beta  = 
\qquad \qquad \qquad \qquad \nonumber \\ 
\int_{-\infty}^{+\infty} \frac{dk\, d^sp}{(2\pi)^{s+1}}\frac{\e^{-ip y_{12}}}{2\omega (k,p)} 
\left \{\frac{\e^{-\beta [\omega(k,p)-\mu]+i\omega (k,p)t_{12}} + \e^{-i\omega (k,p)t_{12}}}
{ 1-\e^{-\beta [\omega(k,p)-\mu]}} \right \} E(k;x_1,x_2;\gamma) \,  .  
\label{htw20}
\eea 
One can easily derive at this point the energy and the charge densities 
\be 
\E (x, \beta; \gamma ) = \langle \theta_{00}(t,x,y) \rangle_\beta \, , \qquad 
 \varrho (x, \beta; \gamma ) = \langle j_0(t,x,y) \rangle_\beta\, , 
\label{densities} 
\ee 
which are both $t$ and $y$-independent because of the symmetry properties 
of the Gibbs state. Setting $\mu = 0$, one finds 
\be  
\E_s (x, \beta; \gamma ) = 
\int_{-\infty}^{+\infty} \frac{dk\, d^sp}{(2\pi)^{s+1}}
\frac{\omega (k,p) \e^{-\beta \omega(k,p)}} 
{ 1-\e^{-\beta \omega(k,p)}} \left \{1+\left [\theta(x) R_+^+(k) + 
\theta(-x) R_-^-(k) \right ] \e^{2 i k x}\right \}, 
\label{hd}
\ee 
and 
\be 
\varrho_s(x,\beta;\gamma ) = 
\int_{-\infty}^{+\infty} \frac{dk\, d^sp}{(2\pi)^{s+1}}
\frac{\e^{-\beta \omega(k,p)}} 
{ 1-\e^{-\beta \omega(k,p)}}\left \{1+\left [\theta(x) R_+^+(k) + 
\theta(-x) R_-^-(k) \right ] \e^{2 i k x}\right \} \, .  
\label{hcd}
\ee 
Since $\omega(k,p)$ depends actually on $|p|$, 
the integration over the angular variables in $p$-space is easily performed 
and gives the area of the unit sphere in $s$ dimensions.  In this way one is left 
with the integrals in $k$ and $|p|$. 

The case of major interest is $s=2$, which corresponds to a plane-defect in 
3+1 space-time dimensions. The identity in the curly brackets 
under the integral (\ref{hd}) precisely reproduces the S-B energy 
\be
\varepsilon_{{}_{\rm S-B}}(\beta )|_{{}_{s=2,\, m=0}} = \frac{\pi^2}{30 \beta^4}  
\ee 
of the free boson field in 3+1 space-time dimensions. 
The correction generated by the $\delta$-impurity reads 
\be 
\varepsilon_2 (x,\beta;\gamma_\eta^+) = \int_0^{+\infty} dk \int_0^{+\infty} d|p|  
\frac{|p| \sqrt{k^2+|p|^2}}{\e^{\beta\sqrt{k^2+|p|^2}}-1}
\frac{[\eta k \sin(2k|x|) -\eta^2 \cos(2k|x|)]}{2\pi^2(k^2 + \eta^2)} \, .   
\label{he}
\ee 
\begin{figure}[h]
\begin{center}
\includegraphics[width=6cm]{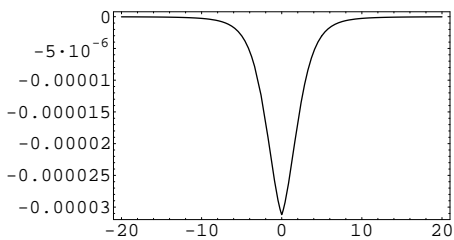}
\end{center}
\caption{Plot of $\varepsilon_2 (x,10;\gamma_{\eta=1})$.} 
\label{f6}
\end{figure} 
It is instructive to compare now the corrections to the S-B law for $s=0$ and 
$s=2$, given by (\ref{eta1}) and (\ref{he}) respectively. 
For this purpose we have plotted $\varepsilon_2$ in Fig. 6 for the same values 
of $\beta$ and $\eta$ for which $\varepsilon$ is displayed 
in Fig. 2. We see that the profiles are the same, but from 
the numerical values on the vertical axes we deduce that 
$\varepsilon_2$ is shifted and squeezed along this axes with 
respect to $\varepsilon$. This phenomenon takes place 
for the charge densities as well. In fact, inserting the data 
(\ref{deltaimp}, \ref{primeimp}, \ref{paritybr}) in the general formula (\ref{hcd}) 
and plotting the resultant expressions, one gets 
Fig.7--9. The comparison between Fig.4--6 and Fig.7--9 
confirms this observation. 
\begin{figure}[h]
\begin{center}
\includegraphics[width=6cm]{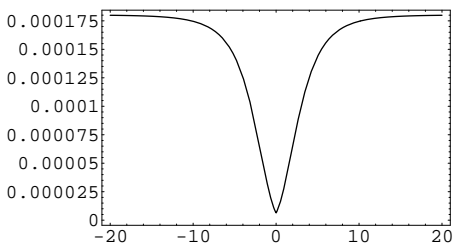}
\end{center}
\caption{Plot of $\varrho_2 (x,10;\gamma_{\eta=1})$ for $m=0.1$.} 
\label{f7}
\end{figure}
\begin{figure}[h]
\begin{center}
\includegraphics[width=6cm]{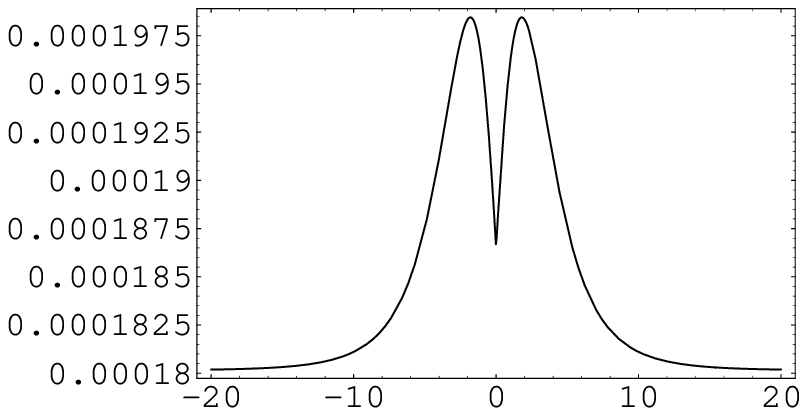}
\end{center}
\caption{Plot of $\varrho_2 (x,10;\gamma_{\xi=1})$ for $m=0.1$.} 
\label{f8}
\end{figure} 
\begin{figure}[h]
\begin{center}
\includegraphics[width=6cm]{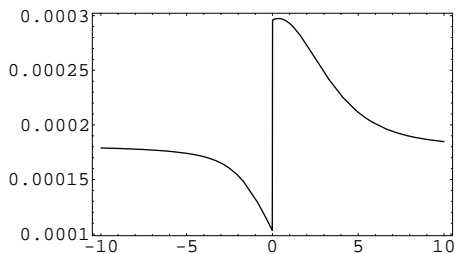}
\end{center}
\caption{Plot of $\varrho_2 (x,10;\gamma_{\zeta=10})$ for $m=0.1$.} 
\label{f9}
\end{figure} 

An aspect which deserves further investigation is the introduction of 
self-interaction, e.g. a term $g\varphi^3$ in the equation of motion (\ref{heqm}). 
The above setup provides the basis for a perturbative investigation of this case. 
Let us mention in this respect that some non-perturbative results about the non-relativistic
$\varphi^4$-theory at zero temperature have been recently obtained 
in \cite{Caudrelier:2004gd, Caudrelier:2004xy}, using the RT algebra technique.   

Another interesting issue concerns the quantum fields 
induced on the defect. Following \cite{Mintchev:2000mf, Mintchev:2001yz}, 
one can show that the limits 
\be 
\lim_{x\to \pm 0} \varphi (t,x,y) = \Phi_{\pm}(t,y) 
\label{induced}
\ee 
exist and correctly define two quantum fields $\Phi_{\pm}(t,y)$, which propagate in the defect. 
$\Phi_+$ and $\Phi_-$ coincide only for parity preserving impurities. 

Summarizing, we have shown in this section that the RT algebra approach 
to impurities has a natural generalization to higher dimensions, where 
impurities have a more realistic physical interpretation. 

\bigskip 

\sect{Outlook and conclusions} 

We studied above the family of impurities defined by all possible self-adjoint extensions 
of the operator $-\partial_x^2$ on functions in $\RR \setminus \{0\}$, showing that 
they can be described in a purely algebraic way. The relevant structure is 
an appropriate RT algebra, which translates in simple algebraic terms the 
solution of the analytic problem defined by eqs. (\ref{eqm}, \ref{initial}, \ref{bc}). 
Constructing the Gibbs state over this algebra, we were 
able to investigate the finite temperature behavior of the associated 
physical systems. Motivated by potential applications in condensed 
matter physics,  we derived in this framework the energy and charge densities 
of the systems in the Gibbs state, discussing the dependence of 
these quantities on the impurity type. We also computed 
in this context the correction to the Stefan-Boltzmann law, corresponding to  
a generic impurity from the above family. 
The contribution of the impurity bound states was discussed as well. 
Self-interactions can be treated in perturbation theory 
with the propagator determined by eq. (\ref{htw20}). 

We started our investigation with point-like defects in 1+1 space-time dimensions, 
demonstrating later on that the RT algebra framework actually applies to 
$s$-dimensional hyperplane-defects in $(s+1)+1$ dimensions for any 
$s\geq 0$.  Since our hyperplane-defects both reflect and transmit, 
they generalize the concept of brane, which is usually assumed to 
reflect only. A challenging open question is if strings give raise to 
such more general configurations. Another interesting issue is 
the study of the fields induced according to (\ref{induced}) 
on impurities and their role in the construction of quantum field theories 
localized on defects. 

RT algebras emerged in the context of factorized scattering theory for  
integrable models in 1+1 dimensions. The results of this paper show that 
they have actually a wider radius of application, representing a efficient 
tool for dealing with impurities at finite temperatures and in higher space-time 
dimensions. 

\bigskip 
%\vfill\eject 


\begin{thebibliography}{99} 

\bibitem{A1} S. Albeverio, F. Gesztesy, R. Hoegh-Krohn ans H. Holden,
Solvable models
in quantum mechanics (Springer-Verlag, Berlin, 1988). 

\bibitem{A2} S. Albeverio, L. Dabrowski and P. Kurasov, 
Lett. Math. Phys. {\bf 45} (1998) 33. 

%\cite{Coutinho:1999xj}
\bibitem{Coutinho:1999xj}
F.~A.~B.~Coutinho, Y.~Nogami and L.~Tomio, 
J.\ Phys.\ A {\bf 32} (1999) 4931. 
[arXiv:quant-ph/9903098].
%%CITATION = QUANT-PH 9903098;%%  

\bibitem{CNP} F.~A.~B.~Coutinho, Y.~Nogami and J.~F.~Perez, 
J.\ Phys.\ A {\bf 30} (1997) 3937. 

\bibitem{SCH} A. G. M. Schmidt, B. K. Cheng and M. G. E. da Luz, 
Phys. Rev. A {\bf 66} (2002) 062712. 

%\cite{Delfino:1994nr}
\bibitem{Delfino:1994nr}
G.~Delfino, G.~Mussardo and P.~Simonetti,
Nucl.\ Phys.\ B {\bf 432} (1994) 518
[arXiv:hep-th/9409076].
%%CITATION = HEP-TH 9409076;%%

%\cite{Konik:1997gx}
\bibitem{Konik:1997gx}
R.~Konik and A.~LeClair,
Nucl.\ Phys.\ B {\bf 538} (1999) 587
[arXiv:hep-th/9703085].
%%CITATION = HEP-TH 9703085;%%

%\cite{Saleur:1998hq}
\bibitem{Saleur:1998hq}
H.~Saleur,
``Lectures on Non-perturbative field theory and quantum impurity  problems",
arXiv:cond-mat/9812110.
%%CITATION = COND-MAT 9812110;%%

%\cite{Saleur:2000gp}
\bibitem{Saleur:2000gp}
H.~Saleur,
``Lectures on Non-perturbative field theory and quantum impurity  problems II'',
arXiv:cond-mat/0007309.
%%CITATION = COND-MAT 0007309;%%

%\cite{Castro-Alvaredo:2002dj}
\bibitem{Castro-Alvaredo:2002dj}
O.~Castro-Alvaredo and A.~Fring,
Nucl.\ Phys.\ B {\bf 649} (2003) 449
[arXiv:hep-th/0205076].
%%CITATION = HEP-TH 0205076;%% 

%\cite{Bowcock:2003dr}
\bibitem{Bowcock:2003dr}
P.~Bowcock, E.~Corrigan and C.~Zambon,
``Classically integrable field theories with defects'',
arXiv:hep-th/0305022.
%%CITATION = HEP-TH 0305022;%%

%\cite{Bowcock:2004my}
\bibitem{Bowcock:2004my}
P.~Bowcock, E.~Corrigan and C.~Zambon,
``Affine Toda field theories with defects'', 
arXiv:hep-th/0401020.
%%CITATION = HEP-TH 0401020;%%

\bibitem{Langmann} M. Halln\"as, E. Langmann, 
``Exact solutions of two complementary 1D quantum
many-body systems on the half-line", arXiv:math-ph/0404023. 

%\cite{Mintchev:2002zd}
\bibitem{Mintchev:2002zd}
M.~Mintchev, E.~Ragoucy and P.~Sorba,
Phys.\ Lett.\ B {\bf 547} (2002) 313.
[arXiv:hep-th/0209052].
%%CITATION = HEP-TH 0209052;%% 

%\cite{Mintchev:2003ue}
\bibitem{Mintchev:2003ue}
M.~Mintchev, E.~Ragoucy and P.~Sorba, 
J.\ Phys.\ A {\bf 36} (2003) 10407
[arXiv:hep-th/0303187].
%%CITATION = HEP-TH 0303187;%%

%\cite{Mintchev:2003kh}
\bibitem{Mintchev:2003kh}
M.~Mintchev and E.~Ragoucy, 
J.\ Phys.\ A {\bf 37} (2004) 425
[arXiv:math.qa/0306084].
%%CITATION = MATH-QA 0306084;%%

\bibitem{BR} O. Bratteli and D. W. Robinson, Operator Algebras and Quantum
Statistical Mechanics, Vol. 2 (Springer-Verlag, Berlin, 1996). 

\bibitem{PBM} A. P. Prudnikov, Yu.  A. Brychkov and O. I. Marichev, Integrals and Series, 
Vol. 1 (Gordon and Breach, New York, 1986).  

%\cite{Caudrelier:2004gd}
\bibitem{Caudrelier:2004gd}
V.~Caudrelier, M.~Mintchev and E.~Ragoucy,
``The quantum non-linear Schr\"odinger model with point-like defect'', 
arXiv:hep-th/0404144.
%%CITATION = HEP-TH 0404144;%% 

%\cite{Caudrelier:2004xy}
\bibitem{Caudrelier:2004xy}
V.~Caudrelier, M.~Mintchev and E.~Ragoucy,
``Solving the quantum non-linear Schrodinger equation with delta-type impurity'',
arXiv:math-ph/0404047.
%%CITATION = MATH-PH 0404047;%% 

%\cite{Mintchev:2000mf}
\bibitem{Mintchev:2000mf}
M.~Mintchev and L.~Pilo, 
Nucl.\ Phys.\ B {\bf 592} (2001) 219
[arXiv:hep-th/0007002].
%%CITATION = HEP-TH 0007002;%%

%\cite{Mintchev:2001yz}
\bibitem{Mintchev:2001yz}
M.~Mintchev, 
Class.\ Quant.\ Grav.\  {\bf 18} (2001) 4801
[arXiv:hep-th/0103259].
%%CITATION = HEP-TH 0103259;%%


\end{thebibliography}
\end{document}